\documentclass{article}
\usepackage{graphicx}
\usepackage{color}

\usepackage{alltt}
\usepackage{amssymb,amsmath,graphicx,enumerate,subcaption,tikz,url,booktabs}
\usepackage[bmargin=0.75in, tmargin =0.75in,lmargin = 0.5in,rmargin = 0.5in]{geometry}
\graphicspath{{manuscript_figure/}}
\usepackage{hyperref}
\hypersetup{colorlinks=true, citecolor=black, urlcolor=cyan}

\usepackage{natbib}
\usepackage{bbold}

\newcommand{\CVE}{\textrm{CVE}}
\newcommand{\MSE}{\textrm{MSE}}

\providecommand{\Tr}{^{\scriptscriptstyle\top}}

\newcommand{\lam}{\lambda}

\title{Cross Validation Approaches for Penalized Cox Regression}
\author{Biyue Dai\\Department of Biostatistics\\University of Iowa
  \and
  Patrick Breheny\\Department of Biostatistics\\University of Iowa}
\date{\today}
\IfFileExists{upquote.sty}{\usepackage{upquote}}{}
\begin{document}
\maketitle

\section*{Abstract}
Cross validation is commonly used for selecting tuning parameters in penalized regression, but its use in penalized Cox regression models has received relatively little attention in the literature. Due to its partial likelihood construction, carrying out cross validation for Cox models is not straightforward, and there are several potential approaches for implementation. Here, we propose two new cross-validation methods for Cox regression and compare them to approaches that have been proposed elsewhere.
Our proposed approach of cross-validating the linear predictors seems to offer an attractive balance of performance and numerical stability. We illustrate these advantages using simulated data as well as using them to analyze data from a high-dimensional study of survival in lung cancer patients.
\section{Introduction}

\par Since its original proposal \citep{Cox1972}, Cox proportional hazards regression has become the most common regression approach for analyzing survival data.  Cox regression utilizes a partial likelihood construction under an assumption of proportional hazards to estimate the regression coefficients without having to specify the underlying baseline hazard.  The ability to avoid choosing a specific parametric distribution for the survival time is very attractive, as time-to-event data are often poorly described by fully parametric models.

This semiparametric flexibility, however, comes at a cost. The Cox regression model can estimate relative risks, but without estimating the baseline hazard, it does not predict the absolute failure time for any given individual.  This poses a challenge to assessing the predictive accuracy of a given Cox regression model.  The challenge is particularly relevant for penalized Cox regression \citep{Tibshirani1997,Fan2002}, as the assessment of predictive accuracy via cross-validation is the standard method for selecting the regularization parameter and deciding upon a model.

\par Standard cross-validation involves dividing the data into $K$ folds, then fitting the model on $K-1$ of those folds (the training set) and assessing prediction accuracy on the remaining fold (the testing set). This process is then repeated, with each fold serving as the testing set exactly once. For Cox regression, the estimated coefficients allow us to quantify the risk for each subject in the training set relative to other members of the training set, but it is not obvious how to use those estimates to quantify the model's accuracy in the testing set.

\par One basic approach is to simply calculate the partial likelihood based on the observations in the test set as a measure for the model's predictive accuracy.  In doing so, we calculate the risk for each member of the test set relative to other members of the test set.  One drawback to this approach is that it becomes unstable when the size of the the test set is small.  In particular, it cannot be applied to leave-one-out cross-validation (LOOCV), as we must have at least two observations in the test set to compare their risk relative to each other.

\par To overcome this drawback, an alternative approach was proposed by \citet{Verweij1993}.  Their approach, which we describe in detail in Section~\ref{Sec:cox-cv-existing}, stabilizes the cross-validated log likelihood, enabling its use even when the number of subjects in each fold is small.  This approach has been widely used and is the default approach to cross-validation in many software programs, such as the R package {\tt glmnet} \citep{glmnet}.  Although this approach fixes the stability issue, we demonstrate here that in practice, it tends to behave conservatively in terms of model selection.

\par In this paper, we propose two alternative ways to carry out cross validation for Cox regression. Instead of applying cross-validation to the partial likelihood directly, we propose applying cross validation to either the linear predictors of the regression model or to the deviance residuals \citep{Therneau1990}. Through simulation studies, we compare these proposed methods with existing cross-validation approaches for LASSO penalized Cox regression in both low- and high-dimensional settings.  
The linear predictor approach offers the best combination of performance and stability, and we recommend using it for regularization parameter selection in penalized Cox models.  We conclude by applying both proposed and existing approaches to a high-dimensional study of survival in lung cancer patients.

\section{Methods}
\label{Sec:methods}

\par In the Cox model, the hazard function for subject $i$ is given by 
\begin{equation*}
  h_{i}(t) = h_{0}(t) \exp( X_{i}\Tr\beta),
\end{equation*} 
where $h_{0}$ is the baseline hazard and $\exp(X_i\Tr\beta)$ is the risk for a subject with covariates $X_i$ relative to that baseline.  The estimation of the coefficients is obtained by maximizing a partial likelihood.  Letting $t_i$ denote the time on study for subject $i$ and $\delta_{i}$ indicate whether or not an event is observed for subject $i$, each subject's individual contribution to that partial likelihood is
\begin{equation}
  \label{eq:cox-pl-subj}
  L_{i}(\beta) = \left \{\frac{\exp ( X_{i}\Tr\beta)}{\sum_{ k \in R(t_{i})}\exp ( X_{k}\Tr \beta)}\right \}^{\delta_{i}},
\end{equation}
where $R(t_{i})$ denotes the set of subjects at risk at time $t_{i}$.  The partial likelihood for the entire sample of $n$ subjects is then given by:
\begin{equation}
  \label{eq:cox-pl-sum}
  L(\beta) =\prod_{i = 1}^{n} L_{i}(\beta).
\end{equation}
Cox regression can be extended by introducing a penalty into the partial likelihood.  In penalized Cox regression, $\beta$ coefficient estimates are obtained by minimizing the objective function
\begin{equation}
  \label{eq:obj}
  Q(\beta) = - \frac{1}{n} \log L(\beta) + P_{\lambda}(\beta),
\end{equation}
where $P_{\lambda}(\beta)$ is a penalty function that depends on a regularization parameter $\lam$.

\par In this paper, we focus upon the LASSO penalty $P_{\lambda}(\beta) = \lambda \sum_{j} |\beta_{j}|$, although the methods we analyze can be used with any penalty as well as with model selection in unpenalized Cox regression. LASSO-penalized cox regression is particularly useful for high dimensional data where the number of covariates $p > n$, such as predicting overall survival in cancer patients based on genome-wide expression measurements. LASSO estimates yield a sparse model where some coefficients are exactly zero. This sparsity pattern changes as we vary $\lam$: at large values of $\lambda$, most or all of the coefficients are 0, but as $\lam$ decreases, more covariates are selected. Selecting $\lambda$ is critical to LASSO estimation in the sense that the model's accuracy suffers if $\lam$ is either too large or too small.

\par
\begin{figure}
  \centering
  \includegraphics[height= 12cm ]{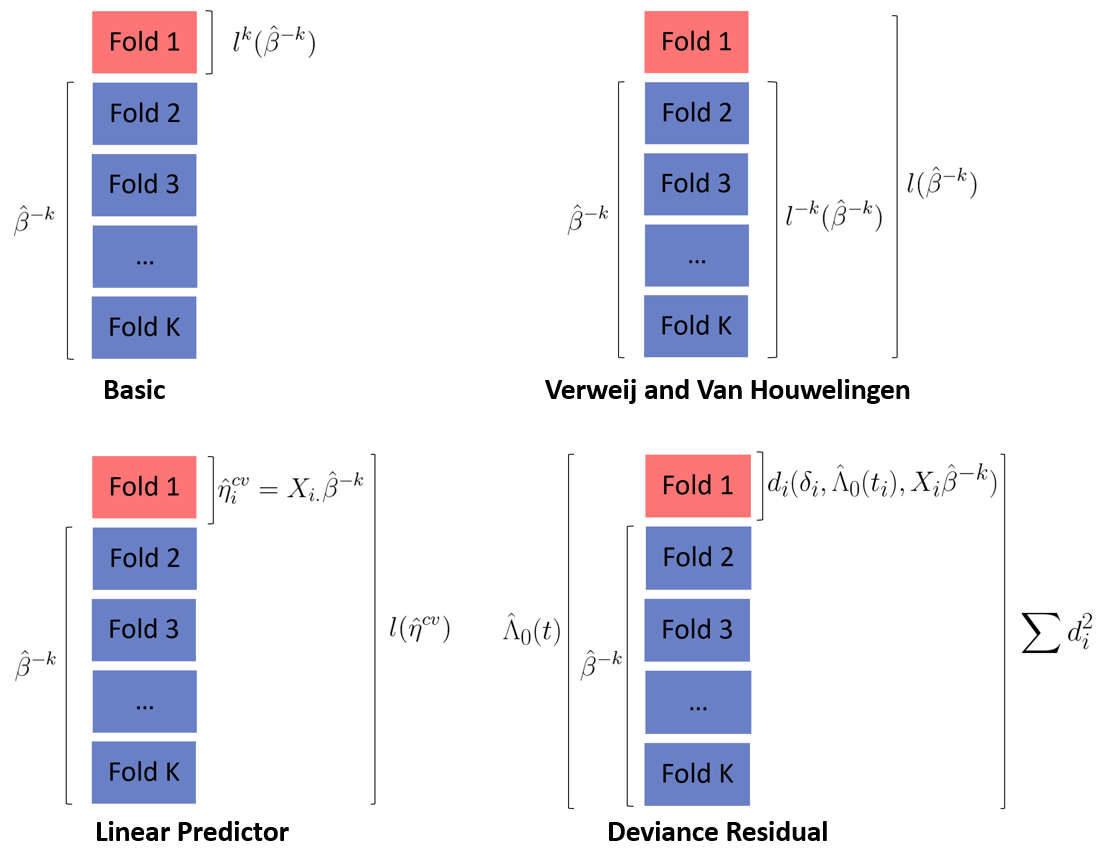}
  \caption{\label{Fig:methods} Illustration of $K$-fold cross validation methods for Cox regression.}
\end{figure}

 Cross-validation is the most common method for selecting $\lambda$ in penalized regression models. Suppose a data set $D$ of n observations is partitioned into $K$ folds: $D_{1}, D_{2}, \ldots, D_{K}$. For a given $k \in \left\{1,2,\ldots, K\right\}$, let $T_{k} = D - D_{k}$ denote the training set. This manuscript is concerned with the question: how should one use the $kth$ fold $D_{k}$ to measure predictive accuracy based on partial likelihood? In particular, the partial likelihood \eqref{eq:cox-pl-subj} involves a risk set -- which observations should be used to form that risk set?  In the following sections, let $L^{-k}$ denote the partial likelihood built using data in $T_k$, let $L^{k}$ denote the partial likelihood built using data in $D_k$, and $L$ denote the partial likelihood built using the entire data set $D$.  The log partial likelihood is denoted by $\ell$, with $\ell^k$ and $\ell^{-k}$ defined similarly. Finally, let $\hat{\beta}^{-k}$ denote the penalized estimates that are obtained by using $L^{-k}(\beta)$ in \eqref{eq:obj}.  Figure~\ref{Fig:methods} diagrams the relationships between the four cross-validation methods we will describe in this section.

\subsection{Cross validated partial likelihood} 
\label{Sec:cox-cv-existing}

\par A direct approach to measure a model's predictive accuracy in $D_k$ is to evaluate the log partial likelihood at $\hat{\beta}^{-k}$ using the data in $D_k$. Cross-validation error (CVE) is then measured by the total deviance across all $K$ folds, which (up to a constant) is given by:
\begin{equation}
  \label{eq:basic}
  \CVE = -2 \sum_{k=1}^{K} \ell^{k}(\hat{\beta}^{-k});
\end{equation}
the factor of 2 is arbitrary, but often used so that CVE estimates the deviance of the model.
We refer this approach as \emph{basic cross-validated partial likelihood}. This is implemented in the $\tt{glmnet}$ package as the \verb|"ungrouped"| option.

\par The basic approach is by far the most common way of conducting cross-validation for models such as linear regression and logistic regression. However, the basic approach can be problematic for Cox regression in the sense that that there may not be enough observations in $D_k$ to build up the risk set for partial likelihood. For example, the basic approach cannot work with leave-one-out cross-validation, since for all folds, $\ell_k$ would either be zero or undefined.

To address this issue, \citet{Verweij1993} proposed an alternative method for cross-validation in Cox models:
\begin{equation}
\label{eq:VVH}
	\CVE = -2\sum_{k = 1}^K \left\{ \ell(\hat{\beta}^{- k})  - \ell^{-k}(\hat{\beta}^{- k}) \right\}. 
\end{equation}
Here, $\ell(\hat{\beta}^{-k})$ is the log partial likelihood evaluated at $\hat{\beta}^{-k}$ using the entire data set $D$ and $\ell^{-k}(\hat{\beta}^{-k})$ is the log partial likelihood evaluated at $\hat{\beta}^{-k}$ on
the training set $T_k$. By avoiding the construction of a partial likelihood on the test set, this approach ensures that the risk set is always sufficiently large and the partial likelihood well-defined.  We refer to this cross-validation method as the Verweij and Van Houwelingen (V\&VH) approach. For many models, such as logistic regression or linear regression, the quantities $\ell(\hat{\beta}^{- k})  - \ell^{-k}(\hat{\beta}^{- k})$ and  $\ell^{k}(\hat{\beta}^{-k})$ are equivalent to each other -- in other words, the basic and V\&VH approaches agree.  However, Verweij and Van Houwelingen's approach is more stable for Cox regression as there is always a large number of observations in the risk sets its calculations are based on.  Since its proposal, the V\&VH approach has been widely implemented as a tool for cross-validation is Cox models.  For example, it is used by the R packages {\tt CoxRidge} \citep{CoxRidge}, {\tt fastcox} \citep{fastcox}, {\tt SGL} \citep{SGL} , {\tt CoxBoost} \citep{CoxBoost}, {\tt mboost} \citep{mboost}, and {\tt glmnet} \citep{glmnet}.  In {\tt glmnet}, it is the default choice for penalized Cox regression and is referred to as the \verb|"grouped"| option in the package's syntax.

\subsection{Cross Validated Linear Predictors}
\label{Sec:linear-predictor}

\par Both of the methods in Section~\ref{Sec:cox-cv-existing} consist of adding together log partial likelihood measures from each fold.  An alternative method is to obtain linear predictors for each fold, then combine these linear predictors to calculate a partial likelihood. We refer to this method as the \emph{cross-validated linear predictors} approach. To be more specific, we would first obtain $\hat{\beta}^{-k}$ from training set $T_{k}$; then the cross-validated linear predictors can be calculated for each observation $i$ from the test set $D_k$:  
\begin{equation}
  \label{eq:cv-lp}
  \hat{\eta}^{cv}_{i} = X_{i}\Tr\hat{\beta}^{-k}
\end{equation} 
After repeating this for all K folds, a complete set of cross-validated linear predictors $\hat{\eta}^{cv} = ( \hat{\eta}^{cv}_{1},  \hat{\eta}^{cv}_{2} , ...  \hat{\eta}^{cv}_{n})$ for the whole sample is obtained. A partial likelihood can then be built using this set of linear predictors to evaluate the predictive accuracy of the model: 
	\begin{equation} 
	L(\hat{\eta}^{cv}) = \prod_{i=1}^{n} \frac{\exp (\hat{\eta}^{cv}_{i})}{\sum_{ j \in R(t_{i})}\exp (\hat{\eta}^{cv}_{j})}.
	\end{equation}
We define the cross-validation error evaluated with cross-validated linear predictors to be $$\CVE = - 2 \log L(\hat{\eta}^{cv}).$$
  
\par This idea of obtaining linear predictors using test sets, then constructing the partial likelihood after pooling all linear predictors together is implemented in R package \texttt{ncvreg} \citep{ncvreg}. This idea has also appeared in the cross-validation literature for quantities that cannot be evaluated on only a subset of the data, such as AUC in logistic and Cox regression models \citep{Parker2007,Simon2011a,Subramanian2011}.

\subsection{Cross Validated Deviance Residuals}

The fundamental challenge of conducting cross validation for Cox regression is that the baseline hazard is not estimated from the model. Hence, another approach to cross-validation would be to include an extra step of estimating the baseline hazard and using it to quantify the cross-validation error. The standard approach to estimating the cumulative baseline hazard, $\Lambda_0$, is the Nelson-Aalen Estimator \citep{nelson1969, aalen1978}:
\begin{equation}
  \hat{\Lambda}_{0}(t) = \sum_{t_j \leq t}\frac{\text{number of failures at time } t_j}{\text{number at risk right before time }t_j}.
\end{equation}
The Cox model implies the following relationship between the baseline cumulative hazard and the cumulative hazard for individual $i$:
\begin{equation}
  \hat{\Lambda}_{i}(t) =  \hat{\Lambda}_{0}(t)\exp(X_i\Tr\hat\beta).
\end{equation}

The baseline hazard could be incorporated into cross-validation in a variety of ways.  Here, we consider the following approach: for fold $k$, we obtain $\hat{\beta}^{-k}$ from the training set $T_k$ and then for each $i$ in the test set $D_k$, 
\begin{equation}
 \label{eq:cv-bhz}
  \hat{\Lambda}^{cv}_{i}(t) =  \hat{\Lambda}_{0}(t)\exp(X_i\Tr\hat\beta^{-k}).
\end{equation}
Note that this uses an estimate of the baseline hazard coming from the full data under the null model, but cross-validation has been applied to the estimation of the regression coefficients $\hat{\beta}$; we will discuss the rationale for this estimator shortly.

A natural candidate for incorporating the cumulative hazard into a loss function is the deviance residual, a normalized form of the Martingale residual \citep{Therneau1990}. First proposed for model diagnostics, deviance residuals have also been occasionally used as a predictive accuracy measure \citep{Therneau2018}.  Given \eqref{eq:cv-bhz}, we can first obtain the cross-validated Martingale residual: 
\begin{equation}
  \hat{M^{cv}_{i}} = \delta_{i} - \hat{\Lambda}_{0}(t_{i})\exp(X_i\Tr\hat\beta^{-k}).
\end{equation}
The cross-validated deviance residuals can then be derived from the Martingale residuals: 
\begin{equation} 
  d_{i} = \text{sgn}(\hat{M}^{cv}_{i})\sqrt{-2(\hat{M}^{cv}_{i} + \delta_{i}\log(\delta_{i} - \hat{M}^{cv}_{i}))}.
\end{equation}
The sum of squared cross-validated deviance residuals, $\sum_{i}\hat{d}_{i}^2$, are then used as the cross validated error, analogously to using the residual sum of squares in linear regression. We refer to this method as the \emph{cross-validated deviance residuals} approach.

The baseline hazard estimate we describe here differs from the one typically used in constructing deviance residuals \citep{Therneau1990}. Instead of the Nelson-Aalen estimator, deviance residuals are typically calculated based on a baseline hazard estimate that has been adjusted for the covariates in the model \citep{breslow1972}. In the context of cross-validation, however, this approach is problematic. Deviance residuals measure the difference between the fitted model and a saturated model; in Cox regression, this saturated model depends on the baseline hazard. Thus, a covariate-adjusted baseline hazard would mean that each fold is compared to a different saturated model. For this reason, it is important that the baseline hazard remains constant across folds when calculating cross-validated deviance residuals; this intuition is borne out by simulations involving various other possible ways of constructing cross-validated deviance residuals (see Supporting Information).  Note that this is not the case for the Brier Score or Kullback Leibler Score (described in Section~\ref{Sec:accuracy}), which involve absolute loss functions, not comparisons with a saturated model.

\section{Simulation Studies}

Simulation studies were conducted to compare how the methods described in Section~\ref{Sec:methods} behave relative to each other in selecting the tuning parameter $\lambda$ and the corresponding model for LASSO penalized regression. For these simulation studies, the covariate matrix $X$, consisting of $n$ observations and $p$ covariates, was generated from independent $\text{Normal}(0, 1)$ samples. Given $X$ and a coefficient vector $\beta$, survival times were generated from exponential distribution with rate parameter $\exp(X\beta)$. Censoring status indicators were drawn independently from the binomial distribution. All simulations were implemented in R \citep{R}. 
   
\subsection {Simulations Comparing Predictive Accuracy}
\label{Sec:accuracy}

In this section, we consider the accuracy of the cross-validation approaches introduced in Section~\ref{Sec:methods}. We varied the sample size, dimension, censoring percentage, and signal strength of the simulated data sets to examine how these factors affect the performance of cross-validation. Accuracy was assesed via several criteria.

Estimation accuracy was measured using the mean squared error of the coefficients ($\MSE$). Suppose $\hat{\beta}(\lambda)$ is the vector of coefficients estimated by a Cox model with LASSO penalty at $\lambda$, and $\beta^*$ the value of $\beta$ used the generate the data. Mean squared error is defined to be 
\begin{equation}
\MSE = E\| \hat{\beta}(\lambda) - \beta^* \| ^2.
\end{equation}

Throughout, we communicate the estimation accuracy in terms of relative MSE, as compared to the oracle model.  For each generated data set, an oracle model was fit using Cox regression only including the variables with non-zero coefficients. Letting $\hat{\beta}^{oracle}$ denote the resulting estimates, the relative MSE, on the log scale, is given by
\begin{equation}
\log\left\{ \frac{\MSE(\hat{\beta}(\lambda^{\text{cv}}))}{\MSE(\hat{\beta}^{\text{oracle}})} \right\},
\end{equation}
where $\lambda^{cv}$ denotes the value of $\lambda$ selected by cross-validation. MSE is estimated by averaging over $N$ replications respectively for $\hat{\beta}(\lambda^{cv})$ and $\hat{\beta}^{oracle}$, with the ratio taken afterwards. The $\log$ is taken to correct for skewness.

\par We also measure the predictive accuracy via the mean squared prediction error, or Brier score \citep{VanHouwelingen2011}. Let $\hat{S}(t_0|\lambda,x)$ denote the predicted probability for an individual with covariates $x$ to survive beyond time $t_0$, and let $y = \mathbb{1}\left\{ t > t_{0}\right\}$ indicate whether the individual actually survived beyond time $t_0$. Then the Brier score is defined to be
\begin{equation}
\text{Brier}(y, \hat{S}(t_0|\lambda,x)) = (y - \hat{S}(t_0|\lambda,x))^2.
\end{equation}
In our simulation study, for each data set, we generated another independent data set with 1000 uncensored observations. We computed Brier scores for all of the cross-validation methods for all individuals in this test set, setting $t_0$ to the median survival time.

\par The third measure we used is the Kullback-Leibler score, which measures the log-likelihood of the prediction model evaluated at the observations \citep{VanHouwelingen2011}. With the same notation defined in the previous paragraph, the Kullback-Leibler score is defined to be
\begin{equation}
	\text{KL}(y, \hat{S}(t_0|\lambda,x)) = -\left\{ y\log(\hat{S}(t_0|\lambda,x) + (1 - y)\log(1 - \hat{S}(t_0|\lambda,x)) \right\}.
\end{equation}
As for the Brier score, KL scores were computed for all CV approaches at the median survival time.  Results were then averaged over the N replications. For both Brier and KL scores, a smaller score represents smaller prediction error, thus indicating better prediction accuracy.

\par The final metric we included in our study was Harrell's C Index \citep{HarrellJr1984}. The C Index is a rank-based statistic that measures the concordance between the linear predictor of the selected model and the observed outcome. Suppose we arbitrarily select two individuals; if the individual with the higher predicted risk also died first, then this would be considered as a concordant pair. The C Index considers all possible such pairs and estimates of proportion of those pairs that are concordant. A C Index of 0.5 means the prediction is no better than flipping a coin and C Index of 1 means perfect concordance. We computed the C Index using the independent test set of 1000 individuals described above.

\begin{figure}[ht]
  \centering
  \includegraphics[height= 7cm ]{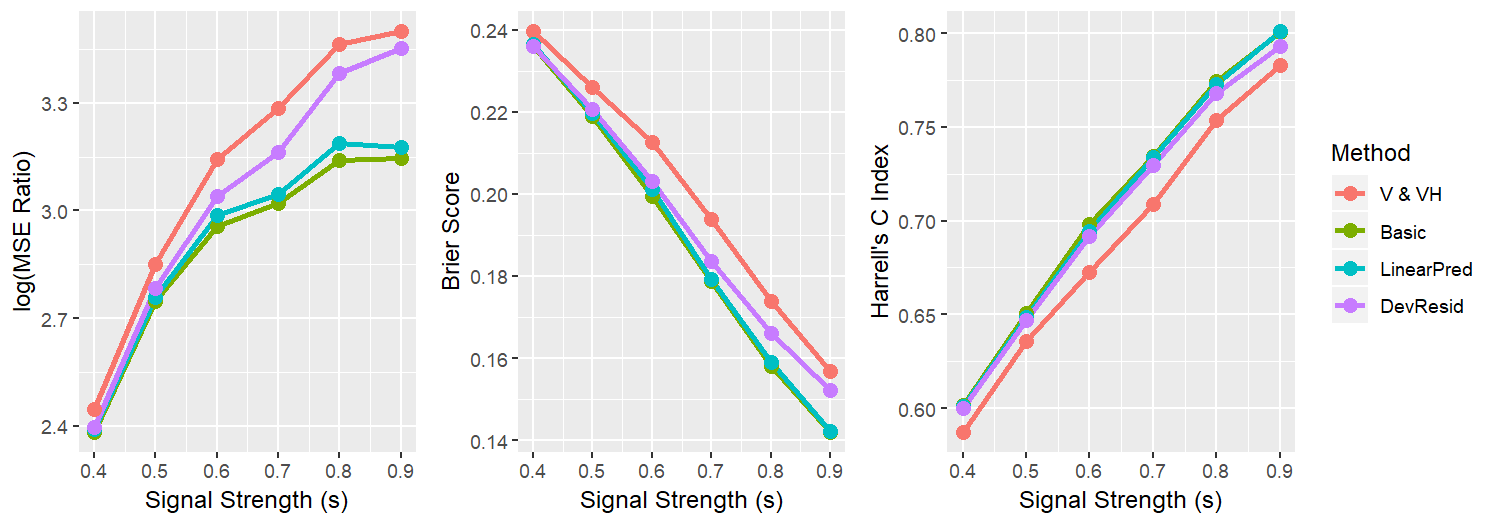}
  \caption{\label{Fig:mse-brier-c} Simulation comparing 4 CV methods. The horizontal axis in all three plots is $s$, which represents the signal strength in the data. The value of $s$ is varied from 0.4 to 0.9.  $\log(\text{MSE ratio})$, relative to the oracle modle is plotted in the left panel. Out-of-sample Brier scores are plotted in the middle. Out-of-sample C index is plotted in the right panel. For each simulated data set, n = 120, p = 1000. Expected censoring percentage is 10$\%$.}
\end{figure}	

For our first simulation study, the covariate matrix has $n = 120$ observations and $p = 1000$ covariates. The covariate vector $\beta$ is assumed to be sparse, with $\beta_j = 0$ for $j \in \{11, 12, \ldots, 1000\}$ and $\beta_j = s$ for $j \in \{1, 2, \ldots, 10 \}$, where $s$ is a prespecified constant that specifies the signal strength in the simulated data set. Here, we vary $s$ from 0.4 to 0.9. The survival outcomes have 10$\%$ expected censoring. For each of the four methods described in Section~\ref{Sec:methods}, 10-fold cross-validation was implemented. $N = 200$ replications were used for each scenario.  Figure~\ref{Fig:mse-brier-c} illustrates the results of this simulation study. For all four metrics, the overall pattern is the same (KL scores not shown in Figure~\ref{Fig:mse-brier-c} due to space constraints, but they were very similar to Brier scores).  Although the four cross-validation methods performed relatively similarly, the basic and linear predictor approaches had the best performance across all metrics: lower MSE, better Brier scores, and higher C Index.

\begin{table}[!htb]

\setlength{\tabcolsep}{3pt}

\caption{\label{Tab:sim} Simulation comparing four cross-validation methods with varying dimension and signal strength}
\centering
\begin{tabular}[t]{cccccccccc}
\toprule
 n& p& $\#$ Non-Zero & Signal & Method & $\lambda$ & log(MSE Ratio) &Brier Score & Kullback Leibler & C Index\\
\midrule
150 & 50 & 5 & weak & V\&VH  & 0.098 (0.030) & 1.561 (0.851) & 0.245 (0.014) & 0.685 (0.030) & 0.610 (0.035) \\
    &    &   &     & Basic  & 0.088 (0.025) & 1.531 (0.838) & 0.244 (0.013) & 0.683 (0.029) & 0.614 (0.029) \\
    &    &   &     & LinearPred  & 0.092 (0.029) & 1.543 (0.846) & 0.245 (0.014) & 0.684 (0.030) & 0.612 (0.034) \\
    &    &   &     & DevResid  & 0.098 (0.029) & 1.547 (0.848) & 0.245 (0.014) & 0.684 (0.030) & 0.611 (0.035) \\
\addlinespace
150 & 50 & 5 & strong & V\&VH  & 0.071 (0.010) & 1.586 (0.989) & 0.176 (0.011) & 0.526 (0.027) & 0.750 (0.012) \\
    &    &   &     & Basic  & 0.066 (0.011) & 1.544 (0.971) & 0.176 (0.011) & 0.526 (0.028) & 0.749 (0.013) \\
    &    &   &     & LinearPred  & 0.067 (0.010) & 1.540 (0.981) & 0.176 (0.011) & 0.525 (0.028) & 0.750 (0.012) \\
    &    &   &     & DevResid  & 0.094 (0.011) & 1.995 (0.841) & 0.181 (0.011) & 0.539 (0.026) & 0.751 (0.013) \\
\addlinespace
400 & 10000 & 20 & weak & V\&VH  & 0.126 (0.019) & 2.958 (0.416) & 0.251 (0.012) & 0.696 (0.026) & 0.589 (0.044) \\
    &    &   &     & Basic  & 0.103 (0.022) & 2.880 (0.437) & 0.244 (0.015) & 0.680 (0.031) & 0.603 (0.039) \\
    &    &   &     & LinearPred  & 0.105 (0.023) & 2.886 (0.437) & 0.244 (0.015) & 0.681 (0.031) & 0.602 (0.039) \\
    &    &   &     & DevResid  & 0.107 (0.022) & 2.895 (0.433) & 0.245 (0.014) & 0.683 (0.03) & 0.602 (0.039) \\
\addlinespace
400 & 10000 & 20 & strong & V\&VH  & 0.084 (0.012) & 3.799 (0.548) & 0.176 (0.03) & 0.528 (0.069) & 0.763 (0.048) \\
    &    &   &     & Basic  & 0.059 (0.007) & 3.455 (0.578) & 0.154 (0.019) & 0.468 (0.049) & 0.780 (0.031) \\
    &    &   &     & LinearPred  & 0.060 (0.007) & 3.479 (0.574) & 0.155 (0.019) & 0.470 (0.049) & 0.780 (0.030) \\
    &    &   &     & DevResid  & 0.076 (0.018) & 3.708 (0.518) & 0.166 (0.024) & 0.502 (0.057) & 0.772 (0.049) \\
\bottomrule
\end{tabular}
\end{table}

Table~\ref{Tab:sim} displays the results of additional simulations carried out as we varied the dimension of the data set along with the signal strength.  Specifically, we examined the following settings: low dimension (p = 50) with weak signal, low dimension with strong signal, high dimension (p = 10000) with weak signal, and high dimension with strong signal. In all settings, the censoring rate was 30\%. For weak cases, signal strength $s$ was set to be 0.3; for strong cases, $s$ was set to be 0.6. 

As in Figure~\ref{Fig:mse-brier-c}, the basic and linear predictor approaches outperform the deviance residual and V\&VH approaches.  The improvement is particularly noticeable in the most difficult setting: high-dimensional data with weak signal.  Table~\ref{Tab:sim} also provides some insight into why the deviance residual and V\&VH approaches perform poorly relative to basic CV and the linear predictor approach: they act conservatively, consistently choosing larger $\lam$ values and thus, selecting fewer variables.

\subsection {Stability}
\label{Sec:stability}

Given that basic cross-validation is both straightforward and demonstrates strong predictive accuracy, as seen in Section~\ref{Sec:accuracy}, one might wonder why it is not more widely used for Cox regression.  The reason is that it suffers from several drawbacks with respect to numerical stability.

First, the number of uncensored observations in the test set may be insufficient to construct a well-defined partial likelihood for basic cross-validation, leaving the overall estimate of cross-validation error undefined as well. When computing the partial likelihood, only individuals whose events are observed will contribute to the likelihood. If there are no events observed in a particular fold, then the partial likelihood for that fold is undefined. 

\begin{figure}[!htb]
  \centering
  \includegraphics[height= 7cm ]{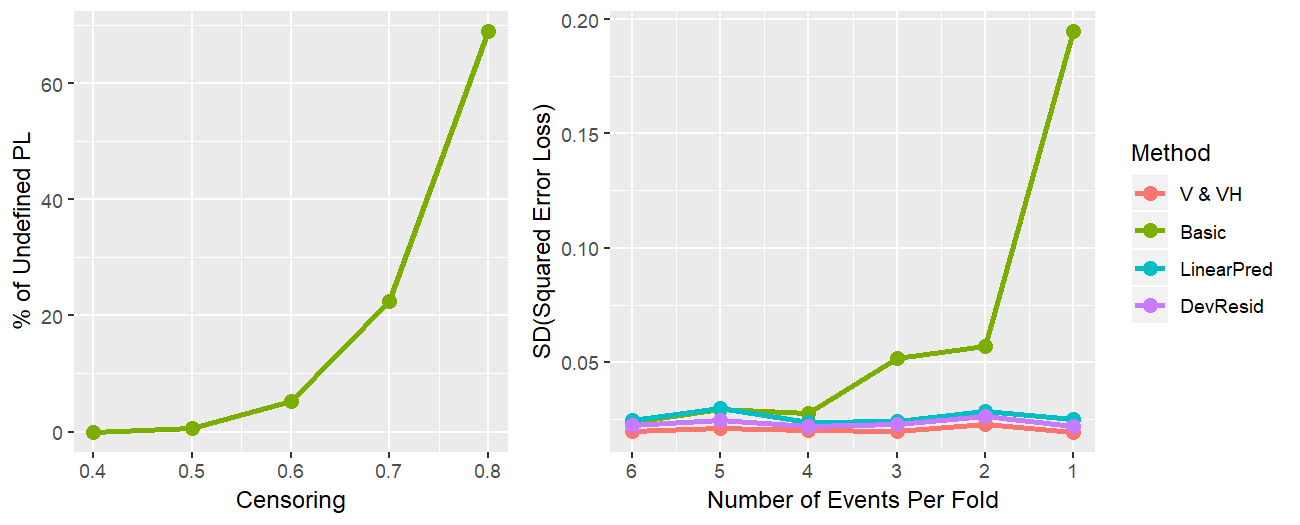}
  \caption{\label{Fig:stability} Simulations illustrating the stability issues with the basic approach. For the figure on the left, $n = 100$, fold assignments are unbalanced, and the vertical axis is the proportion of undefined partial likelihoods. For the figure on the right, $n = 120$, fold assignments are balanced so that the number of events in each fold is the same, and the vertical axis is the standard deviation of the squared error loss. Basic cross-validation is much more variable than other approaches when the number of events in each fold is small.}
\end{figure}

To illustrate how common this problem is, we generated 1000 independent data sets of $n = 100$ observations with a fixed censoring rate. For each data set, we conducted 10-fold cross-validation and recorded whether or not there was at least one event in each fold (i.e., whether the partial likelihood was well-defined for all folds). We repeated the same mechanism for scenarios with different censoring rates, ranging from $40\%$ to $80\%$. As illustrated in the left panel in Figure~\ref{Fig:stability}, at $n=100$ the basic approach can be problematic even at moderate censoring rates. For example, basic CV cannot be used with over 20\% of data sets when $70\%$ of observations are censored.

One approach for alleviating this issue is to balance the censoring status across all folds when making fold assignments. That is, to ensure the number of events observed is the same (or within 1) across all folds. Unless the number of the folds exceeds the number of observed events, such as in leave-one-out cross-validation, this approach will guarantee the partial likelihood in each fold is well-defined for the basic approach, thus the overall cross-validated error is well-defined. 

In this case, however, even though the partial likelihood is well-defined, basic cross-validation suffers from extreme variability.  To illustrate, we generated data sets with $n = 120$ observations and a fixed censoring rate of $50\%$, so that there are 60 observed events. For each data set, we carried out cross-validation using balanced fold assignments with the number of folds set to be 10, 12, 15, 20, 30 and 60. When the number of folds was set to be 10, there were exactly 6 events in each fold; when the number of folds was set to be 60, there was exactly one event in each fold, and so on. After each fold assignment, we computed the cross-validated error using all four approaches described in Section~\ref{Sec:accuracy}. We recorded the model that minimizes the cross-validation error for each approach and computed its squared error loss. We replicated the whole process for $N = 200$ time. In the end, we computed the standard deviation of the squared error loss for each approach. 

The results of this simulation are illustrated in the right panel of Figure~\ref{Fig:stability}. While the other three approaches are not affected by the number of folds and number of events in each fold, the variability of basic cross-validation increases dramatically as the number of events per fold decreases. When there is a sufficient number of events in each fold, the basic approach is similar to the other three approaches. However, when there are only one or two events in each fold, the basic approach can be up to 10 times more variable than other methods, with a corresponding increase in the mean squared error.  In these situations, therefore, even though the basic approach is defined and CVE can be calculated, its performance is very poor.

To shed some light on why the variability of basic cross-validation explodes in the 60-fold scenario, note that in this case, there is exactly one event and one censored observation in each fold. In many folds, the uncensored event happens after the other observation has been censored. In this situation, the risk set at the time immediately prior to the event contains only a single observation. Its contribution to the log partial likelihood is therefore 0, which is well-defined but not informative. This situation can happen regardless of the fold size and censoring rate, but it happens far more often as the number of events per fold decreases.

Based on the simulation results illustrated in this section, we suggest that one should proceed with caution if they want to use the basic approach: (i) it is best to balance the censoring status when creating fold assignments for the basic approach, (ii) avoid using the basic approach in conjunction with a high number of folds, and (iii) avoid using the basic approach when the number of observed events is small. Alternatively, use the linear predictor approach, which has similar performance without the stability issues.

\subsection {Leave-One-Out Cross Validation}
\label{Sec:loocv}

As discussed earlier, basic cross-validation cannot be used to carry out leave-one-out cross validation (LOOCV) for Cox regression models. The other three approaches (V\&VH, linear predictor, and deviance residual), however, are not subject to the stability issues described in Section~\ref{Sec:stability}, and can be used to implement LOOCV.  Here, we compare these three approaches for carrying out LOOCV, and also compare LOOCV to 10-fold CV.

\begin{figure}[!htb]
  \centering
  \includegraphics[height= 8cm ]{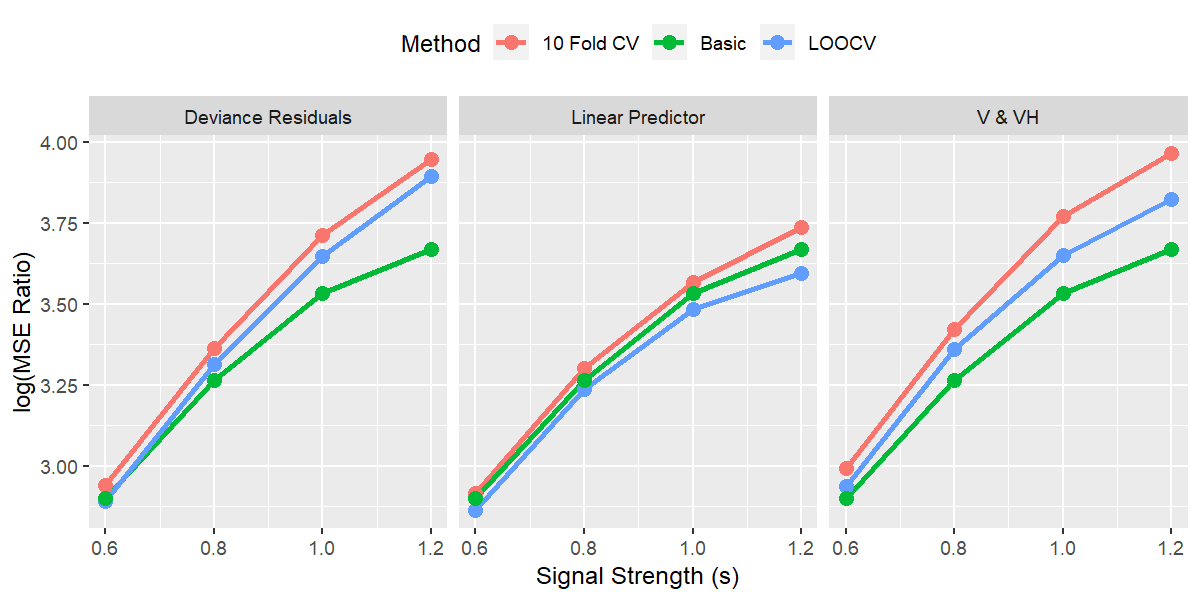}
  \caption{\label{Fig:loocv}Simulation comparing LOO and 10-fold CV across three different approaches. With $N = 200$ replications, the data sets generated have dimension $n = 120$ and $p = 1000$. The vertical axis is the $\log$(MSE ratio). In each panel, the LOO version and 10-fold version of the same approach are plotted. In all three panels, 10-fold basic cross-validation is included as a baseline for comparison (blue line).}
\end{figure}	

For this simulation experiment, we generated datasets with $n = 120$ and $p = 1000$, using the same data generating procedure as described in Section~\ref{Sec:accuracy}. For each of the V\&VH, linear predictor, and deviance residual approaches, we conducted both 10-fold CV and LOOCV. We also conducted basic 10-fold cross-validation as a baseline for comparison. We recorded the model that minimizes the cross-validated error  for each approach. After $N = 200$ replications, $\log$ of the MSE ratio was computed using the same definition described in Section~\ref{Sec:accuracy}.

Results of the simulation are shown in Figure~\ref{Fig:loocv}. The panels illustrate the results for each CV approach, one with the 10-fold version and one with the leave-one-out version. The 10-fold basic approach was plotted in all three panels as a baseline for comparison.  For all three approaches, the Leave-One-Out version selected models with a smaller $\log(\text{MSE ratio})$ than the 10-fold version. For the V\&VH and deviance residual approaches, however, LOOCV was still inferior to basic 10-fold CV.  As described in Section~\ref{Sec:accuracy}, these approaches tend to be conservative, selecting a larger-than-optimal value of the regularization parameter; LOOCV helps with this problem, but does not alleviate it entirely.

On the other hand, the LOOCV version of the linear predictor approach surpasses basic 10-fold CV, with a smaller $\log(\text{MSE ratio})$ across all signal strength levels. At the same time, even the 10-fold version of the linear predictor approach outperforms the leave-one-out version of the V$\&$VH approach and the deviance residual approach. In other words, although adopting a leave-one-out approach improves the predictive accuracy for V$\&$VH and deviance residual cross-validation, the magnitude of improvement in predictive performance is smaller than the improvement gained from switching to the linear predictor approach.

\subsection {Cross-Validated C Index}
\label{Sec:CIndex}

\par All of the cross-validation approaches described in Section~\ref{Sec:methods} use the Cox partial likelihood, in one way or another, as the criterion upon which model selection is based.  In this section, we compare these likelihood-based approaches against an alternative approach of cross validation using concordance, also known as the cross-validated C Index.  This is a widely used approach for conducting cross-validation with time-to-event outcomes, particularly for machine learning models \citep{Subramanian2011,Simon2011a}. This approach typically proceeds by constructing cross-validated linear predictors, as in Section~\ref{Sec:linear-predictor}, but then using those predictors to calculate concordance as opposed to partial likelihood.

\par
\begin{figure}[!htb]
  \centering
  \includegraphics[height= 15cm ]{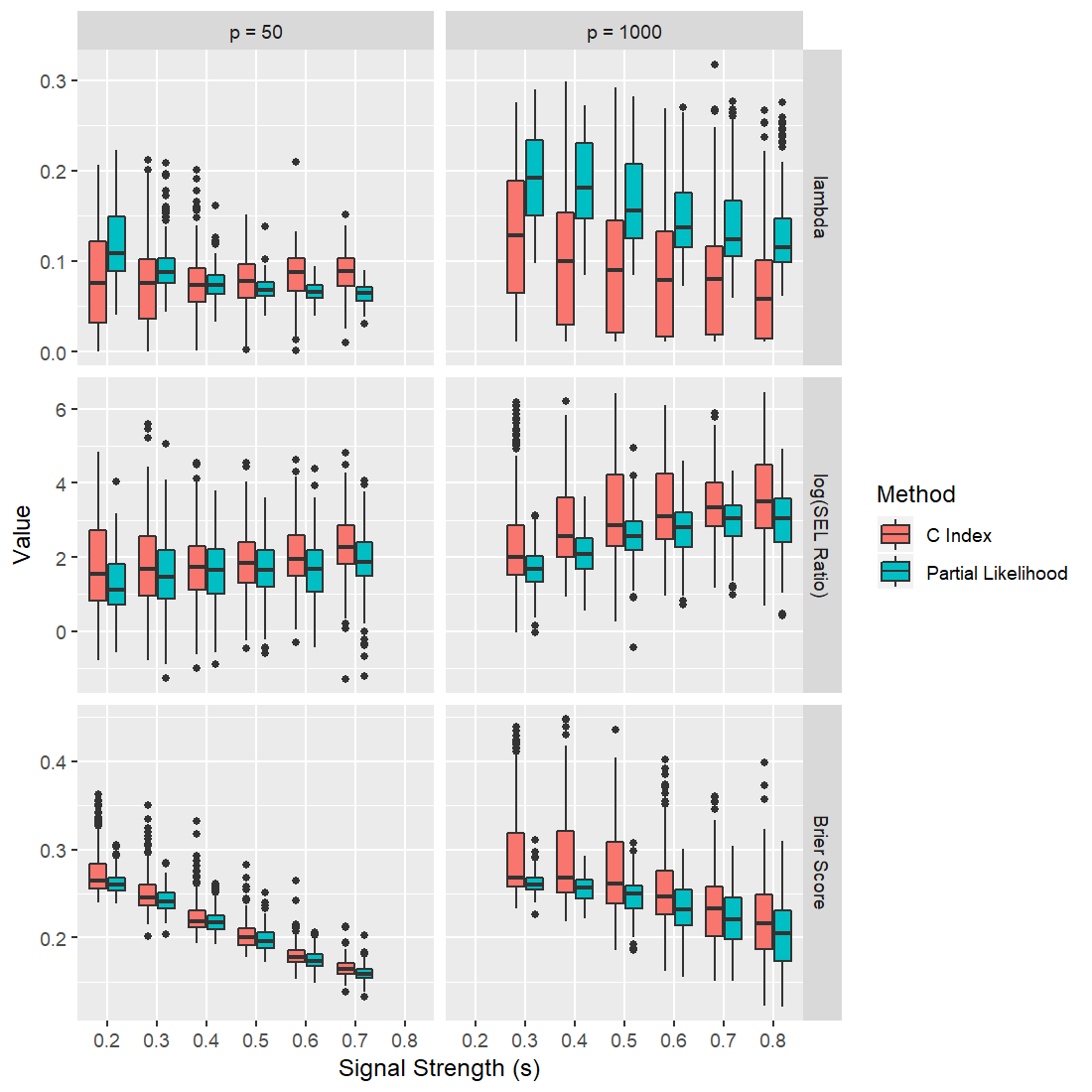}
  \caption{\label{Fig:cvauc}Comparing Partial Likelihood and C Index both built from Cross-validated Linear Predictors. The horizontal axis represents the strength of signal $s$ in the data. The left figures are under the low dimension scenarios where p = 50 and n = 150. The right figures are under the high dimension scenarios where p = 1000 and n = 120. The expected percentages of censoring for both scenarios are 30 $\%$. The top row are box plots of the selected tuning parameter $\lambda$. The mid row plots the $\log(\text{relative Squared Error Loss})$, with respect to the oracle model. The bottom row plots the Brier Scores. }
\end{figure}

 We conducted simulations in both low dimensional ($p = 50$, $n = 150$) and high dimensional ($p = 1000$, $n = 120$) settings. For each scenario, we generated data from a spectrum of signal strength as in Section~\ref{Sec:accuracy}. Results of the simulations are illustrated in Figure~\ref{Fig:cvauc}.  We used the linear predictor approach (CV-LP) here as the representative approach for likelihood-based cross-validation.

Several observations can be drawn from Figure~\ref{Fig:cvauc}. First, likelihood-based CV tends to select more accurate models, as judged by both mean squared error and Brier score.  Second, likelihood-based CV is also more stable than the AUC-based CV, in the sense that it tends to select a narrower range of models, with more consistent performance, than CV-AUC.  Lastly, CV-AUC is consistently more liberal than likelihood-based CV, selecting models with smaller values of the regularization parameter $\lam$, and therefore preferring models with larger coefficient values and more nonzero coefficients.

\section{Application to Real Data}

In this section, we demonstrate the usage and performance of the aforementioned cross-validation methods when fitting penalized regression models on data from a study of gene expression and survival in patients with lung cancer. \citet{shedden2008gene} conducted a large, retrospective study to validate prognostic gene expression signatures as predictors of overall survival in lung cancer patients. They gathered expression profiles of 22,283 genes for 442 non-small-cell lung cancer (NSCLC) patients along with some additional clinical and demographic data. About $50\%$ of the patients died during the course of the study (236 events), with the rest censored. 

For this analysis, we fit elastic net models to the data in addition to lasso models, as these models often improve upon the lasso when many correlated genes are involved \citep{Zou2005}. The clinical variables stage, age and gender were included in the model without penalization, as it is reasonable to assume that all three have an effect on survival. In addition to the regularization parameter $\lam$, the elastic net model has an extra tuning parameter $\alpha$, where $\alpha=1$ corresponds to lasso and $\alpha=0$ corresponds to ridge regression. For $\alpha \in \{0.3, 0.5, 0.7, 0.9, 1\}$, we selected a regularization parameter $\lam$ that has the lowest cross-validation error for each of the four cross-validation methods described in Section~\ref{Sec:methods}.  Although each method can be used to select a pair of $\alpha$ and $\lam$ values, the methods are on different scales and cannot be directly compared with each other.  For this reason, Table~\ref{tb:shedden} reports the cross-validated C-Index and Brier Scores for each method's choice of $\lam$ at each value of $\alpha$.  As shown in the table, the elastic net model with $\lam$ and $\alpha=0.3$ selected by the cross-validated linear predictor approach has the best performance in terms of C Index and Brier Score.

\begin{table}[ht]
\centering
\caption{\label{tb:shedden} The C Index and Brier Score of all elasticnet models selected by cross-validation methods }
\begin{tabular}{lccccccccccc}
\hline
  & \multicolumn{5}{c}{Cross-validated C Index} & \multicolumn{5}{c}{Brier Score} \\ 
  & $\alpha = 0.3$ & $\alpha = 0.5$ & $\alpha = 0.7$ & $\alpha = 0.9$ & $\alpha = 1$ &  
 & $\alpha = 0.3$ & $\alpha = 0.5$ & $\alpha = 0.7$ & $\alpha = 0.9$ & $\alpha = 1$ \\ \cline{2-6} \cline{8-12} 
V $\&$ VH      & 0.639 & 0.637 & 0.638 & 0.637 & 0.637 &  & 0.243 & 0.244 & 0.244 & 0.245 & 0.245 \\
Basic       & 0.647 & 0.639 & 0.638 & 0.638 & 0.639 &  & 0.233 & 0.243 & 0.244 & 0.244 & 0.244 \\
Linear Pred    & \textbf{0.650} & 0.647 & 0.638 & 0.638 & 0.639 &  & \textbf{0.229} & 0.232 & 0.244 & 0.244 & 0.244 \\
Deviance Resid & 0.645 & 0.637 & 0.638 & 0.637 & 0.637 &  & 0.235 & 0.244 & 0.244 & 0.245 & 0.245 \\ \hline
\end{tabular}
\end{table}

\begin{figure}[h]
  \centering
  \includegraphics[height= 8cm ]{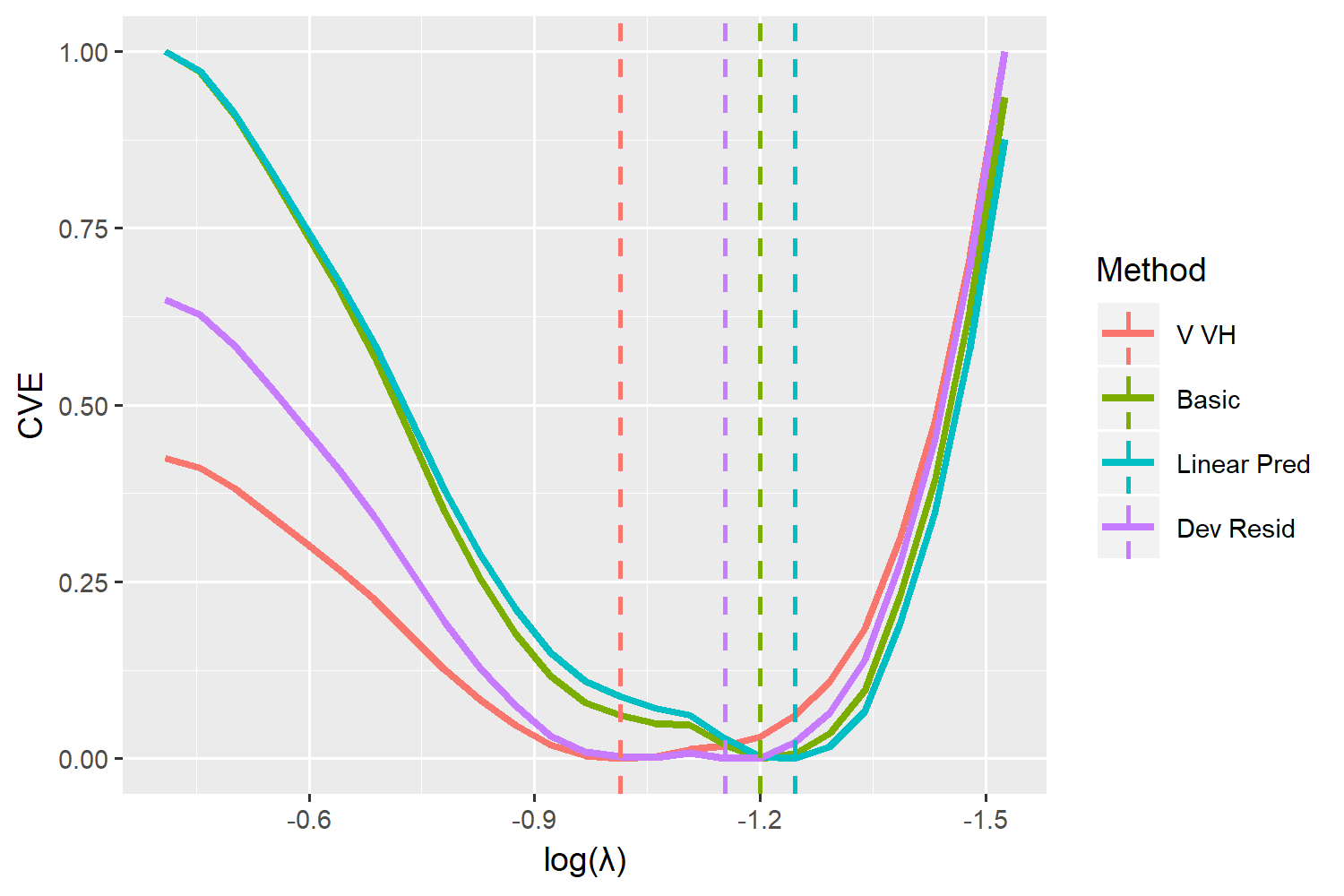}
  \caption{\label{fig:CVE}Rescaled Cross Validated Errors at $\alpha = 0.3$. The horizonal axis is $\log(\lambda)$. The vertical axis is the cross-validated error rescaled to a range from 0 to 1 in order to allow plotting of the four methods on a common axis. The vertical dashed lines represent the $\lambda$ selected by a specific method. The linear predictor approach selected the smallest $\lambda$ and Verweij and Van Houwelingen's approach selected the greatest $\lambda$.}
\end{figure}

For each fixed $\alpha$, we compared the cross-validation errors (CVE) from the four cross-validation methods along the solution path. At $\alpha = 0.3$, CVEs are rescaled and plotted for all four methods along the same grid of $\lambda$ values in Figure~\ref{fig:CVE}. The selected $\lambda$ values, where a given CVE curve reaches its lowest point, are indicated with dotted lines.  The CVE curves for the basic and linear predictor approaches have greater curvature near their minima, and thus offer more clear support for the selected value of $\lam$.  In comparison, the deviance residual and V\&VH approaches are flatter near their minima, supporting a considerably wider range of $\lam$ values and leading to greater model uncertainty.   The selected $\lambda$ values, represented by the vertical dashed lines, and their corresponding models, reflect what was observed in the simulation studies. The linear predictor approach was most liberal and selected 85 genes. While the basic approach and deviance residual approach selected 68 genes and 60 genes respectively, Verweij and Van Houwelingen's approach was most conservative, selecting only 34 genes.  Similar patterns were observed for other $\alpha$ values.

As mentioned earlier, for the elastic net model with tuning parameters chosen by cross-validated linear predictors ($\lam = 0.650$, $\alpha = 0.3$), 85 genes were selected, of which 78 were well-annotated.  We carried out an enrichment analysis of these 78 genes using ToppGene \citep{Chen2009c}; the five most significantly enriched molecular functions are listed in Table~\ref{Tab:go}.  Overall, the results are consistent with our understanding of cancer: many of the genes selected by the elastic net model are involved in signaling and the binding of transcription factors and receptors.  These are precisely the pathways that tend to be disrupted in cancer.  Chloride channels have also recently been implicated in several cancers as therapeutic targets \citep{Peretti2015}.

\begin{table}[ht]
\centering
\caption{\label{Tab:go} Five most significantly enriched molecular functions among genes selected by the linear predictor method with the elastic net ($\alpha=0.3$).  The five terms have a false discovery rate of 3\% using the Benjamini-Hochberg method. Enrichment: frequency of the annotation among the selected genes divided by its frequency in the genome as a whole.}
\begin{tabular}{llrrr}
\toprule
ID & Name & Selected & Enrichment & p \\
\midrule
GO:0008134 & transcription factor binding & 10 & 4.1 & 0.00015 \\
GO:0035257 & nuclear hormone receptor binding & 5 & 7.4 & 0.00057 \\
GO:0019869 & chloride channel inhibitor activity & 2 & 53.2 & 0.00061 \\
GO:0003700 & DNA-binding transcription factor activity & 14 & 2.6 & 0.00068 \\
GO:0005102 & signaling receptor binding & 16 & 2.4 & 0.00083 \\
\bottomrule
\end{tabular}
\end{table}

Of particular interest is the fact that three of the selected genes (PIK3R1, PDPK1, and ITGB1) are involved in the PTEN pathway.  The PTEN gene is an important tumor suppressor; mutations in PTEN are among the most common genetic changes found in human cancers and have been linked to cancers of the lung, breast, prostate, brain, and more.  Time-to-event analyses like this one of gene expression data provide valuable insights for developing gene signatures that can help to identify patients with more severe disruptions to important pathways such as PTEN, thereby informing prognosis and impacting treatment decisions.

\section{Discussion}

Cross-validation is a crutial step in fitting penalized Cox regression models. In this paper, we propose two cross-validation methods for Cox Regression: cross-validated linear predictors and cross-validated deviance residuals. We conducted a series of simulation experiments to study and compare their performance with two existing approaches, basic cross-validation and Verweij and Van Houwelingen's cross-validated partial likelihood, in selecting tuning parameters when fitting penalized regression models.

Despite the fact Verweij and Van Houwelingen's cross-validated partial likelihood has been widely implemented, its predictive performance appears suboptimal, particularly when the dimension of the data is large and/or signals are weak. As illustrated in both simulation studies and real data example, Verweij and Van Houwelingen's approach tends to perform conservatively, selecting fewer features than all other models. In the lung cancer data example, the number of genes selected by other methods almost doubled the number of genes selected by Verweij and Van Houwelingen's approach. From an application standpoint, using Verweij and Van Houwelingen's approach in high-dimensional applications may lead to more false negatives and miss important true findings.

While the basic approach performed well in terms of predictive performance, this approach suffers from a lack of numerical stability as discussed in Section~\ref{Sec:stability}. The cross-validated linear predictor approach performed very similarly to the basic approach in terms of predictive performance, but without any problems due to numerical stability.  We recommend using the proposed linear predictor approach in practice for selecting regularization parameters in penalized Cox regression.

\bibliographystyle{ims-nourl}

\end{document}